# Single crystal study of the layered heavy fermion compounds $Ce_2PdIn_8$, $Ce_3PdIn_{11}$, $Ce_2PtIn_8$ and $Ce_3PtIn_{11}$

M. Kratochvilova[1*], M. Dusek[2], K. Uhlirova[1], A. Rudajevova[1], J. Prokleska[1], B. Vondrackova[1], J. Custers[1], V. Sechovsky[1]

[1] Department of Condensed Matter Physics, Faculty of Mathematics and Physics, Charles University, Prague, Ke Karlovu 5, 121 16, Czech Republic

[2] Department of Structure Analysis, Institute of Physics, Prague, Cukrovarnicka 10, 162 00, Czech Republic

We report on single crystal growth and crystallographic parameters results of $Ce_2PdIn_8$, $Ce_3PdIn_{11}$, $Ce_2PtIn_8$ and $Ce_3PtIn_{11}$. The Pt-systems $Ce_2PtIn_8$ and $Ce_3PtIn_{11}$ are synthesized for the first time. All these compounds are member of the $Ce_nT_mIn_{3n+2m}$ ($n$ = 1, 2,..; $m$ = 1, 2,.. and $T$ = transition metal) to which the extensively studied heavy fermion superconductor $CeCoIn_5$ belongs. Single crystals have been grown by In self-flux method. Differential scanning calorimetry studies were used to derive optimal growth conditions. Evidently, the maximum growth conditions for these materials should not exceed 750 °C. Single crystal x-ray data show that $Ce_2TIn_8$ compounds crystallize in the tetragonal $Ho_2CoGa_8$ phase (space group P4/$mmm$) with lattice parameters $a$ =4.6898(3) Å and $c$ =12.1490(8) Å for the Pt-based one (Pd: $a$ = 4.6881(4) Å and $c$ = 12.2031(8) Å). The $Ce_3TIn_{11}$ compounds adopt the $Ce_3PdIn_{11}$ structure with $a$ = 4.6874(4) Å and $c$ = 16.8422(12) Å for the Pt-based one (Pd: $a$ = 4.6896 Å and $c$ = 16.891 Å). Specific heat experiments on $Ce_3PtIn_{11}$ and $Ce_3PdIn_{11}$ have revealed that both compounds undergo two successive magnetic transitions at $T_1$ ~ 2.2 K followed by $T_N$ ~ 2.0 K and $T_1$ ~ 1.7 K and $T_N$ ~ 1.5 K, respectively. Additionally, both compounds exhibit enhanced Sommerfeld coefficients yielding $\gamma_{Pt}$ = 0.300 J/mol K$^2$ Ce ($\gamma_{Pd}$ = 0.290 J/mol K$^2$ Ce), hence qualifying them as heavy fermion materials.

## 1. Introduction

The series of compounds with the chemical formula $Ce_nT_mIn_{3n+2m}$ ($n$ = 1, 2; $m$ = 1 and $T$ = Co, Rh, Ir) have become an intensively investigated group of heavy fermion (HF) materials in which superconductivity and magnetic ordering coexist in a broad range of the temperature-pressure phase diagram [1]. The general formula $Ce_nT_mIn_{3n+2m}$ reflects the tetragonal crystal structures built of $n$ blocks of $CeIn_3$ and $m$ blocks of $TIn_2$ stacking along the $c$-axis. For instance, in $CeTIn_5$, each $CeIn_3$ layer is separated by one $TIn_2$ layer. This configuration leads to a 2D character of the Ce sublattice. Another example is $Ce_2TIn_8$. Here two $CeIn_3$ layers are separated by one $TIn_2$ layer, and the structure dimensionality partially evolves from 2D to 3D [1,2,3]. Recently, new structures and compounds, like $CePt_2In_7$ [4], $Ce_2PdIn_8$ [5, 6], $Ce_5Pd_2In_{19}$ and $Ce_3PdIn_{11}$ [7] with $T$ = Pd or Pt have been discovered and subjected to studies of their crystal structure and physical properties. In conjunction with the already known $Ce_nT_mIn_{3n+2m}$ compounds, they significantly enrich the spectrum of dimensionality evolution from 2D-like to 3D systems. Besides varying pressure [8] and

---

* Corresponding author. Tel.: +420-606-523-841

*E-mail address*: marie@mag.mff.cuni.cz

substitution doping [9], structural dimensionality represents another parameter suitable for tuning the quantum critical point in the $Ce_nT_mIn_{3n+2m}$ compounds, spanning from more 2D-like $CeTIn_5$ (or even $CePt_2In_7$) to more 3D-like $Ce_3PdIn_{11}$ and $Ce_3PtIn_{11}$ structures [2,7,10,11].

Solution growth technique has been proven suitable for obtaining phase pure and of excellent quality single crystals of the $Ce_nTIn_{3n+2}$ ($n$ =1, 2, $T$ = Co, Rh, Ir) systems [1,2,3,8]. However, the situation in the new, Pd- and Pt-based compounds is different. Already the growth of bulk $Ce_2PdIn_8$ single crystals has been found to be difficult; the phase was mostly acquired as a very thin layer (50 - 100 μm) growing on top of $CeIn_3$ single crystals [13,14]. Single phase samples could only be successfully obtained by careful mechanical separation of the two phases [14,15].

Recently two new phases in the $Ce_nPd_mIn_{3n+2m}$ line were discovered. These two phases, $Ce_3PdIn_{11}$ and $Ce_5Pd_2In_{19}$ were obtained in polycrystalline form after annealing the melt at 600 °C [7] together with the already known compound $Ce_2PdIn_8$. By recalling previous studies of isothermal sections of the Ce-Pd-In system performed at 500 °C [5], 600 °C [7] and 750 °C [16] which showed that the $Ce_2PdIn_8$ compound exists only in the lower two temperatures, it can be inferred that these more complex structures are unstable at temperatures above ~ 750 °C.

In this article, we report on single crystal growth of new phases $Ce_2PtIn_8$ and $Ce_3PtIn_{11}$ [10], and on recently reported compounds $Ce_3PdIn_{11}$ [7, 10] and $Ce_2PdIn_8$ [5, 6]. The temperature ranges for the crystal growth have been optimized based on results of experiments utilizing differential scanning calorimetry (DSC). Detailed structure characterization of the prepared single crystals will be presented.

**2. Experimental**

Single crystalline samples were grown by self-flux method [14]. Appropriate starting compositions of high purity elements (Ce 99.9 % further purified by solid state electrotransport [17], Pd 99.995 %, Pt 99.995 %, In 99.999 %) with total weight of 5 g were placed in alumina crucibles. Another crucible, filled with quartz wool was placed on top of it. The quartz wool acted as a filter to separate the solid crystals from the remaining flux when decanting directly after the growth process. The single crystals were grown from starting compositions and in the temperature ranges listed in Table 1.

Differential thermal analysis (DTA) has proven to be very helpful tool for determining the growth conditions for solution growth method of single crystals [18]. We performed differential scanning calorimetry (DSC) and DTA study of Ce-Pd-In system utilizing a SETSYS Evolution 24 instrument (SETARAM Instrumentation). Standard alumina crucibles with diameter 0.5 cm and height 0.8 cm were used; the total mass of each sample was ~ 100 mg. The temperature dependence of the heat flow was measured in He atmosphere and the heating/cooling rate was 10 K/min. The transition temperatures were determined as onset of the observed peaks. Ce-Pd-In systems were heated up to various temperatures between 350 °C and 1000 °C. In all performed experiments, we focused on the first thermal cycle starting with pure elements in molar ratio 2:1:25. After each thermal process, the products were carefully analyzed by microprobe analysis. The single crystals grown during the DSC experiment were very small. Hence, we omitted further investigation techniques such as single crystal diffraction or microprobe analysis of polished crystals in order to disclose inclusions of other phases.

Except the microprobe analysis, the interpretation of the data was based also on the knowledge of Ce-In and Pd-In binary phase diagrams [19,20]. Some events in the studied Ce-Pd-In system were ascribed to formation of binary phases and to confirm that, DTA measurements of Ce-In and Pd-In binary systems with the same Ce:In and Pd:In molar ratios as in the Ce-Pd-In systems were performed.

Microprobe analysis was realized using Scanning Electron Microscope (SEM) Tescan Mira LMH equipped with the energy dispersive x-ray detector (EDX) Bruker AXS. We used a point analysis and elemental mapping option. As discussed previously [13,14], the backscattered electrons (BSE) do not provide any contrast to resolve $CeIn_3$ and $Ce_2PdIn_8$ from each other. Therefore, elemental mapping option was mostly used to characterize sample homogeneity and to disclose inclusions of neighboring $Ce_nT_mIn_{3n+2m}$ phases. By employing long acquisition times, the elemental mapping shows clear contrast between the $Ce_nT_mIn_{3n+2m}$ phases. However, in a sense of quantitative composition analysis, the difference between neighboring phases such as $Ce_3PdIn_{11}$ and $Ce_5Pd_2In_{19}$ or $Ce_2PdIn_8$ and $Ce_5Pd_2In_{19}$ is close to the resolution limit of 1-2 % (absolute). The final phase identification was done by single crystal x-ray diffraction.

The single crystal diffraction on selected samples was performed using x-ray diffractometer Gemini, equipped with Mo lamp with graphite monochromator and Mo-Enhance collimator producing Mo $K_\alpha$ radiation, and CCD detector Atlas. The samples exhibited diffraction patterns with rather long reflection profiles due to large mosaicity. Nevertheless they could be easily solved and refined due to the simplicity of their structure. We used programs Superflip [21] for structure solution and Jana2006 [22] for structure refinement based on $F^2$. Very important for obtaining reliable results was the absorption correction based on the crystal shape in combination with the empirical correction based on spherical harmonic functions. Both corrections were done by the data processing software CrysAlis [23].

Specific heat ($T > 0.45$ K) was measured by a hybrid adiabatic relaxation method using a Quantum Design PPMS 9 T equipped with the Helium-3 system.

### 3. Results and discussion

*3.1. Differential scanning calorimetry and crystal growth*

The binary Ce-In and Pd-In systems in molar composition 2:25 and 1:25, respectively, were studied by DTA up to temperature of 1000 °C. In Fig. 1a the first thermal cycle for the Ce-In system is shown. First, the heating shows an endothermic peak at ~156 °C in agreement with the melting point of indium. The next peak with onset at ~270 °C is related to the highly exothermic reaction of Ce with In leading to formation of mostly solid Ce-In binary compounds. These compounds decompose in a broad temperature range around ~ 900°C [19], which manifest in a broad endothermic bump between 800-950 °C. Upon cooling, a peak with onset at 900 °C is observed. The peak position is in agreement with the liquidus line reported for Ce-In binary diagram [19] and marks the solidification of $CeIn_3$; no other events were observed down to solidification temperature of indium. In Fig. 1b, the Pd-In binary system is shown. After the melting of indium, a broad exothermic bump with a maximum around 580 $^{o}$C was found representing the formation of Pd-In binary phases. The cooling curve shows a sharp peak at 460 $^{o}$C which, consistently with the Pd-In phase diagram [20], indicates the solidification of $Pd_3In_7$. No other events occur down to solidification of indium.

The ternary Ce-Pd-In system with Ce:Pd:In molar ratio 2:1:25 was studied by DSC up to various temperatures. In Fig. 2a, a heating and cooling cycle up to 750 $^{o}$C is shown. The heating branch of the temperature dependence of the heat flow shows first the consumption of heat during melting of In. It is succeeded by an exothermic event at ~280 $^{o}$C, which indicates the formation of the Ce-In binary compounds similar to the Ce-In binary system (Fig. 1a). It is followed by broad bump with maximum at ~600 $^{o}$C. There reactions of the Ce-In system with Pd take place. Upon cooling, an exothermic event at ~550 $^{o}$C is observed which we ascribe to the formation of $Ce_2PdIn_8$ and $Ce_3PdIn_{11}$ ternary phases. Microprobe analysis of the products confirmed the creation of these compounds and only spurious presence of $CeIn_3$ in the batch after this DSC treatment. Fig. 3a shows the elemental mapping of a selected area of the sample subjected to this DSC experiment. The marked $Ce_2PdIn_8$ compound was identified by microprobe analysis.

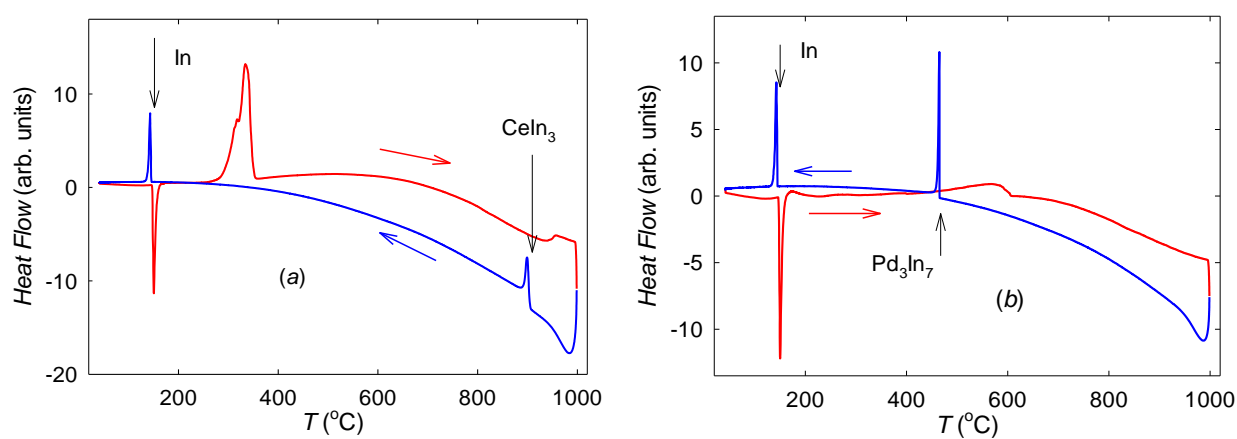

**Fig. 1**. DTA curves for the binary Ce-In (*a*) and Pd-In system (*b*) measured up to 1000 °C. The red and blue lines represent heating and cooling processes, respectively (also marked by red and blue arrows). While cooling, crystallization of $CeIn_3$ (*a*), $Pd_3In_7$ (*b*) and solidification of In (marked by arrows) was detected.

In Fig. 2b, the DSC analysis of the Ce-Pd-In system performed up to 1000 $^{o}$C is depicted. This experiment simulates some of the previous $Ce_2PdIn_8$ single crystal growth experiments [6,13] and sheds light on the thermal stability of the ternary compounds. Up to 750 $^{o}$C, the heating curve reproduces the previous experiments (Fig. 2a). When heating further up to the maximum temperature of 1000 $^{o}$C, a broad endothermic event around 900 $^{o}$C is detected. It is related to decomposition of phases formed at lower temperatures. Upon cooling, an exothermic peak at ~900 $^{o}$C is observed. Consistently with Fig. 1a, we ascribe this peak to crystallization of $CeIn_3$. Below 570 $^{o}$C, small features are noticeable. With utmost probability, these peaks belong to crystallization of the ternary phases $Ce_2PdIn_8$ and $Ce_3PdIn_{11}$. At 430 $^{o}$C, a sharp exothermic peak corresponding to $Pd_3In_7$ crystallization emerges. The presence of the compounds $Ce_2PdIn_8$, $Ce_3PdIn_{11}$, $CeIn_3$ and $Pd_3In_7$ in the sample was confirmed by microprobe analysis with a majority of the latter two. This result is consistent with the high temperature growth process [6,13].

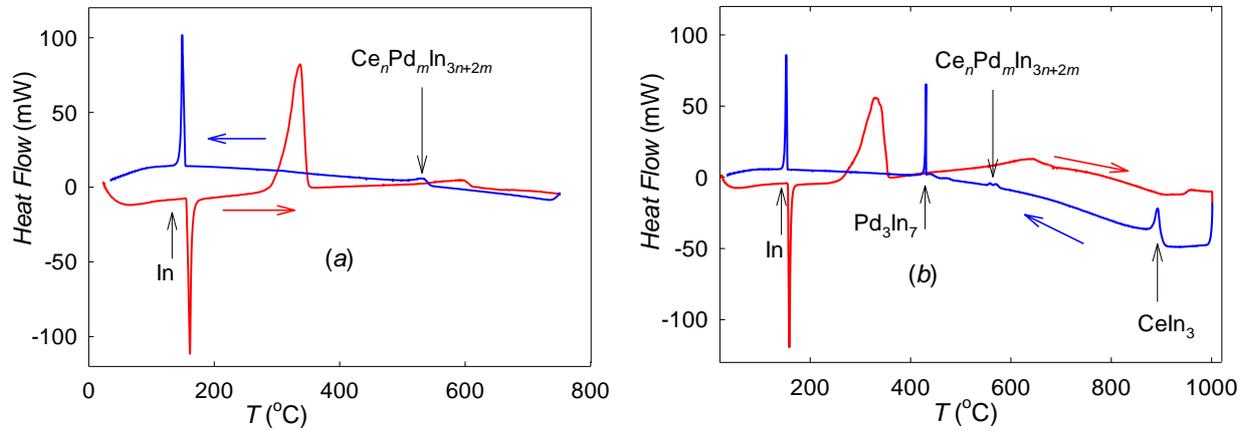

**Fig. 2**. DSC curves for the ternary Ce-Pd-In system measured up to 750 °C (*a*) and up to 1000 °C (*b*). The red and blue lines represent heating and cooling processes, respectively (also marked by red and blue arrows).

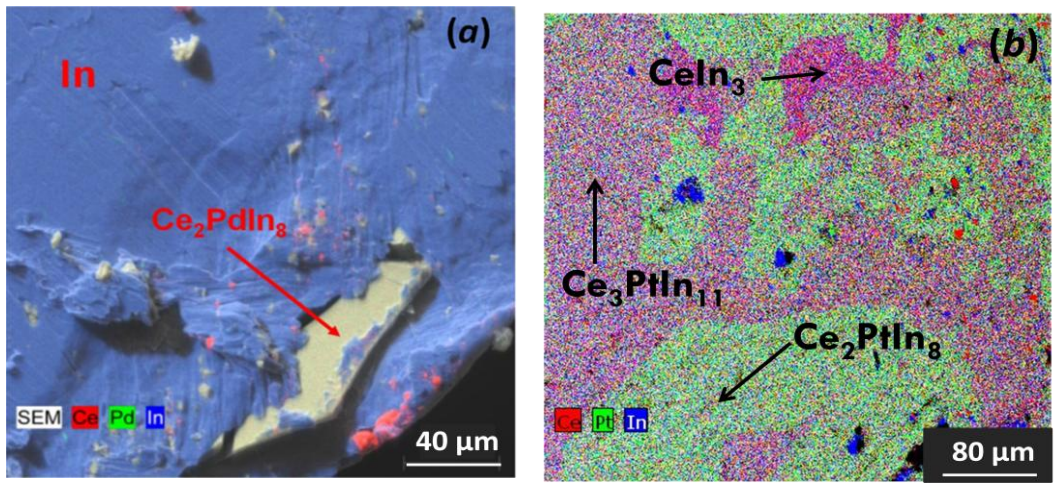

**Fig. 3.** Elemental mapping of a cut through the sample subjected to DSC heating and cooling cycle up to 750 °C (*a*). Elemental mapping of the polished section of a sample taken from batch marked by a dagger (†) in Tab 1 (*b*). Areas marked by arrows were identified as $CeIn_3$, $Ce_2PtIn_8$ and $Ce_3PtIn_{11}$.

Based on the results from the DSC analysis, the single crystal growth was realized at temperatures below 750 °C. Because the kinetics of the single crystal growth process cannot be fully reproduced by the DCS (with a cooling rate two orders of magnitude faster), several growth attempts were performed in order to tune the yield of the desired phases. The range of the growth temperatures and starting composition together with the main growth products are summarized in Table 1. The cooling rate during the growth was 1-3°C/h. Decreasing the temperature below 750 °C suppressed the formation of $CeIn_3$ significantly; single crystals of of $Ce_2PdIn_8$ and $Ce_3PdIn_{11}$ were found to be the majority products. The $Ce_3PdIn_{11}$ crystals were almost an order of magnitude larger (up to 7 mg) and seem to grow at higher temperatures than $Ce_2PdIn_8$; decanting the flux at higher temperatures led to better phase homogeneity of $Ce_3PdIn_{11}$, while the low-temperature growth (starting from 450°C) favoured the formation of $Ce_2PdIn_8$. Still, the yield of $Ce_2PdIn_8$ was rather low and the crystals contained $Ce_3PdIn_{11}$ inclusions (often hidden inside the crystals). Besides the various temperature ranges we also tested the effect of starting elemental stoichiometry. The phase homogeneity of $Ce_3PdIn_{11}$ was improved by changing the growth composition from 2:1:25 to Ce-richer system such as 3:1:25 or 5:2:50. The 5:2:50 ratio suggests that $Ce_5Pd_2In_{19}$ could be obtained. However, this phase was not found in any batches.

A similar growth procedure was applied to the Ce-Pt-In system resulting in two new phases, $Ce_3PtIn_{11}$ and $Ce_2PtIn_8$. The growth conditions and products are listed in Table 1. Very similar to the Pd-based compounds, when growing from the starting temperature of 1000 °C, $Ce_2PtIn_8$ and $Ce_3PtIn_{11}$ form a thin layer (<30 µm) on top of $CeIn_3$ single crystals. At lower growth temperatures $Ce_3PtIn_{11}$ forms rather easily, contrary to $Ce_2PtIn_8$ which is difficult to isolate without $Ce_3PtIn_{11}$ impurities. The character of the impurities present in the multiphase samples is illustrated by elemental mapping image shown in Fig. 3b. $Ce_2PtIn_8$ single-phase crystals with dimensions < 0.2 mm were isolated after a long inspection of the growth products by microprobe and single crystal x-ray analysis.

**Table 1**
Growth conditions for the $Ce_nT_mIn_{3n+2m}$ compounds. The starting (decanting) temperature is marked $T_1$ ($T_2$). In the table, we only list the most important or majority phases. Spurious amounts of $CeIn_3$ were detected in all growth batches.

| $T$ | Ce:$T$:In | $T_1/T_2$(°C) | Final products |
|---|---|---|---|
| Pd | 2:1:25 | 950/350 | $Ce_2PdIn_8$<50µm layer, $CeIn_3$ [13,14] |
| Pd | 2:1: 25 | 750/300 | $Ce_2PdIn_8$, $Ce_3PdIn_{11}$ |
| Pd | 3:1:25 | 750/550 | $Ce_3PdIn_{11}$ |
| Pd | 5:2:50*‡ | 750/550 | $Ce_3PdIn_{11}$ |
| Pt | 2: 1: 25 | 1000/400 | $Ce_2PtIn_8$<30µm layer, $CeIn_3$ |
| Pt | 2:1: 25*† | 750/300 | $Ce_2PtIn_8$, $Ce_3PtIn_{11}$, $Ce_3Pt_4In_{13}$ |
| Pt | 3:1:25*‡ | 750/550 | $Ce_3PtIn_{11}$ |
| Pt | 1:2:30 | 1000/400 | $CeIn_3$, $CePt_2In_7$, $Ce_3Pt_4In_{13}$ |
| Pt | 1:2:30 | 750/300 | $CePt_2In_7$, $Ce_3Pt_4In_{13}$ |

* Structure parameters presented in Tables 2-4 were obtained on samples from these batches.
‡ Samples from these batches were used for the specific heat measurements.
† A sample from this batch is shown in Fig. 2b.

We have tested the growth conditions for $CePt_2In_7$ single crystals reported in Ref. [11]. Starting from the molar composition 1:2:30 and the growth temperature range 1000-400 °C, single crystals of $CePt_2In_7$ were obtained. However, more than 50% of the analysed crystals from the batches were $CeIn_3$ or $Ce_3Pt_4In_{13}$ [25]. Lowering the growth temperature suppressed the $CeIn_3$ formation but did not significantly influence the $CePt_2In_7$:$Ce_3Pt_4In_{13}$ yield ratio. Therefore, also in this case a careful analysis of the crystals is necessary prior to further studies.

*3.2. Single crystal x-ray diffraction*

Selected samples of $Ce_2PdIn_8$, $Ce_3PdIn_{11}$, $Ce_2PtIn_8$ and $Ce_3PtIn_{11}$ were subject to single crystal x-ray diffraction. Results on $Ce_2PdIn_8$ were in agreement with previous studies [7,14] and are $a = 4.6881(4)$ Å, $c = 12.2031(8)$ Å. The $Ce_3PdIn_{11}$ structure investigation has been reported only very recently and because it was performed on single crystals isolated from polycrystalline samples [7], we include the analysis of this compound for comparison. The structure parameters of $Ce_3PdIn_{11}$ obtained from our study are in agreement with the reported ones. The new phases $Ce_2PtIn_8$ and $Ce_3PtIn_{11}$ crystallize in the $Ho_2CoGa_8$- and $Ce_3PdIn_{11}$-type structures, respectively [7,26]. The structural parameters of $Ce_3PdIn_{11}$, $Ce_2PtIn_8$ and $Ce_3PtIn_{11}$ are presented in Table 2 and 3; the interatomic distances are shown in Table 4. The data were obtained from samples selected from the growth-batches marked in Table 1 by an asterisk.

**Table 2**
Crystallographic data for the compounds $Ce_3PdIn_{11}$, $Ce_3PtIn_{11}$ and $Ce_2PtIn_8$.

| Compound | $Ce_3PdIn_{11}$ | $Ce_3PtIn_{11}$ | $Ce_2PtIn_8$ |
|---|---|---|---|
| Lattice parameters (Å) (single-crystal data) | $a = 4.6896(11)$ $c = 16.891(3)$ | $a = 4.6874(4)$ $c = 16.8422(12)$ | $a = 4.6898(3)$ $c = 12.1490(8)$ |
| Cell volume (Å$^3$) | 371.47(14) | 372.44(9) | 267.21(3) |
| Formula weight (g.mol$^{-1}$) | 1789.74 | 1878.4 | 1393.84 |
| Density (g.cm$^{-3}$) | 7.998 | 8.427 | 8.682 |
| $\mu$(mm$^{-1}$) | 26.81 | 35.14 | 38.29 |
| Crystal size (mm) | 0.3 x 0.11 x 0.08 | 0.83 x 0.55 x 0.30 | 0.28 x 0.17 x 0.11 |
| Reflections measured, independent, independent with I>3$\sigma$, $R_{int}$ | 1510, 348, 269, 0.058 | 5150, 361, 348, 0.107 | 3814, 261, 255, 0.047 |
| $R(F^2>3\sigma)$, $wR(F^2)$, S | 0.054, 0.132, 1.63 | 0.037, 0.104, 1.65 | 0.022, 0.076, 1.53 |
| No of parameters, No of restraints | 22, 0 | 22, 0 | 17, 0 |
| Weighting scheme | | $w=1/[\sigma^2(I)+0.0016I^2]$ | |
| Extinction coefficient | 110(20) | 390(40) | 100(17) |
| $\Delta\rho_{max}$, $\Delta\rho_{min}$ (eÅ$^{-3}$) | 6.87, -5.58 | 5.76, -5.32 | 1.32, -2.10 |

**Table 3**
Structure parameters for the compounds $Ce_3PdIn_{11}$, $Ce_3PtIn_{11}$ and $Ce_2PtIn_8$.

| Atom | Wyckoff position | x | y | z | $U_{eq}$ |
|---|---|---|---|---|---|
| $Ce_3PdIn_{11}$ | | | | | |
| Ce1 | 2g | 0 | 0 | 0.27683(8) | 0.0123(4) |
| Ce2 | 1a | 0 | 0 | 0 | 0.0133(5) |
| Pd1 | 1b | 0 | 0 | 0.5 | 0.0156(7) |
| In1 | 4i | 0.5 | 0 | 0.41180(8) | 0.0159(5) |
| In2 | 2h | 0.5 | 0.5 | 0.27797(11) | 0.0173(5) |
| In3 | 4i | 0.5 | 0 | 0.13805(8) | 0.0186(5) |
| In4 | 1c | 0.5 | 0.5 | 0 | 0.0177(7) |
| $Ce_3PtIn_{11}$ | | | | | |
| Ce1 | 2g | 0 | 0 | 0.27755(5) | 0.0093(3) |
| Ce2 | 1a | 0 | 0 | 0 | 0.0096(4) |
| Pt1 | 1b | 0 | 0 | 0.5 | 0.0104(3) |
| In1 | 4i | 0.5 | 0 | 0.41167(6) | 0.0131(3) |
| In2 | 2h | 0.5 | 0.5 | 0.27784(7) | 0.0138(4) |
| In3 | 4i | 0.5 | 0 | 0.13792(5) | 0.0149(4) |
| In4 | 1c | 0.5 | 0.5 | 0 | 0.0136(5) |
| $Ce_2PtIn_8$ | | | | | |
| Ce | 2g | 0 | 0 | 0.30795(7) | 0.0068(3) |
| Pt1 | 1a | 0 | 0 | 0 | 0.0085(2) |
| In1 | 4i | 0 | 0.5 | 0.12241(8) | 0.0109(3) |
| In2 | 2e | 0 | 0.5 | 0.5 | 0.0121(4) |
| In3 | 2h | 0.5 | 0.5 | 0.3072(6) | 0.0117(3) |

**Table 4**
Interatomic distances in $Ce_3PdIn_{11}$, $Ce_3PtIn_{11}$ and $Ce_2PtIn_8$.

| | $Ce_3PdIn_{11}$ | | | $Ce_3PtIn_{11}$ | | | $Ce_2PtIn_8$ | |
|---|---|---|---|---|---|---|---|---|
| Atom | To atom | d (Å) | Atom | To atom | d (Å) | Atom | To atom | d (Å) |
| Ce1 | In1 | 3.2704 | Ce1 | In1 | 3.2551 | Ce1 | In1 | 3.2501 |
| | In2 | 3.3161 | | In2 | 3.3145 | | In2 | 3.3054 |
| | In3 | 3.3156 | | In3 | 3.3200 | | In3 | 3.3127 |
| Ce2 | In3 | 3.3069 | Ce2 | In3 | 3.2998 | | | |
| | In4 | 3.3160 | | In4 | 3.3145 | | | |
| In1 | In1 | 3.3160 | In1 | In1 | 3.3145 | In1 | In1 | 3.3127 |
| | In1 | 2.9800 | | In1 | 2.9755 | | In1 | 2.9730 |
| | In2 | 3.2570 | | In2 | 3.2516 | | In3 | 3.2442 |
| | Pd1 | 2.7780 | | Pt1 | 2.7760 | | Pt1 | 2.7743 |
| In2 | In3 | 3.3291 | In2 | In3 | 3.3235 | In2 | In2 | 3.3127 |
| In3 | In3 | 3.3160 | In3 | In3 | 3.3145 | | In3 | 3.3114 |
| | In4 | 3.3069 | | In4 | 3.2998 | | | |

### 3.3. Specific heat of $Ce_3PdIn_{11}$ and $Ce_3PtIn_{11}$

$Ce_3PdIn_{11}$ and $Ce_3PtIn_{11}$ single crystals of masses $m_{Pd}$ = 8.93 mg and $m_{Pt}$ = 3.44 mg were studied by means of specific heat down to ~ 0.45 K. $Ce_3PdIn_{11}$ shows a double-peak anomaly at $T_1$ ~ 1.7 K and $T_N$ ~ 1.5 K (see Fig. 4), indicating the onset of magnetic order and an order-to-order transition into the final low-temperature magnetic structure, respectively. Compared to results on polycrystalline $Ce_3PdIn_{11}$ [27], the peaks are sharper and shifted to higher temperatures. The specific heat of the $Ce_3PtIn_{11}$ sample reveals similar character to its Pd counterpart with magnetic transitions at $T_1$ ~ 2.2 K and later at $T_N$ ~ 2.0 K.

Since the attempts to synthesize lanthanum counterparts has not been successful so far, we determine the magnitude for the phonon contribution from a $C/T = \gamma + \beta T^2$ fit to the data (fit interval: 6 K $< T <$ 11 K). The values of Sommerfeld coefficient yield $\gamma$ = 290 mJ mol$^{-1}$Ce K$^{-2}$ ($\gamma$ = 300 mJ mol$^{-1}$Ce K$^{-2}$) for $Ce_3PdIn_{11}$ ($Ce_3PtIn_{11}$) qualifying both compounds as HF materials. The values of the $\beta$ coefficient equal to 5.2 mJ mol$^{-1}$Ce K$^{-4}$ (10.9 mJ mol$^{-1}$Ce K$^{-4}$) for $Ce_3PdIn_{11}$ ($Ce_3PtIn_{11}$) and correspond to a Debye temperature $T_D$ = 194 K (201 K).

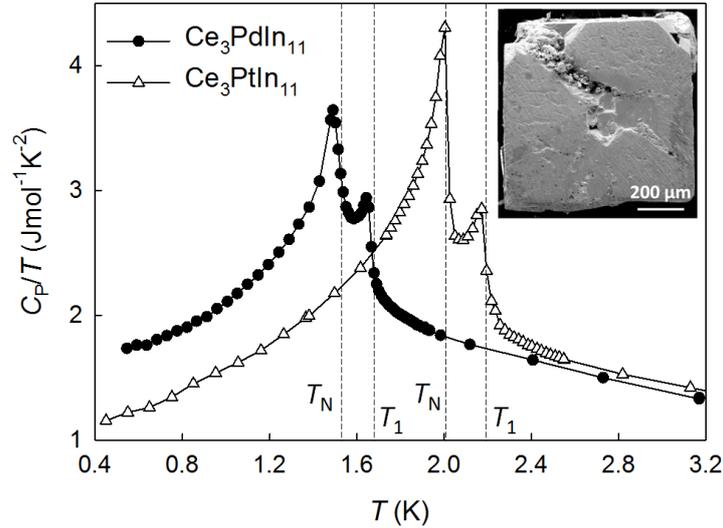

**Fig. 4.** Temperature dependences of the specific heat of $Ce_3PdIn_{11}$ and $Ce_3PtIn_{11}$ single crystals in zero magnetic field. The dashed lines mark the magnetic transitions at $T_1$ and at $T_N$. SEM image of $Ce_3PtIn_{11}$ single crystal of a typical size is shown in the inset.

### 4. Conclusions

Single crystals of the $Ce_3PdIn_{11}$, $Ce_3PtIn_{11}$ and $Ce_2PtIn_8$ compounds have been grown for the first time. Single crystal x-ray diffraction confirmed that the crystal structure of $Ce_3PdIn_{11}$ is in agreement with Ref. [7]. The new phases $Ce_3PtIn_{11}$ and $Ce_2PtIn_8$ crystallize in the $Ce_3PdIn_{11}$-type and $Ho_2CoGa_8$-type structures, respectively.

The solution growth from lower temperatures allows for almost full suppression of $CeIn_3$ formation in both Pd- and Pt-based systems as suggested by the differential scanning calorimetry analysis of the growth process. For the tested growth conditions, $Ce_3PdIn_{11}$ and $Ce_3PtIn_{11}$ were the most stable compounds. Single crystals of the $Ce_3PdIn_{11}$ or $Ce_3PtIn_{11}$ were present in a majority of the batches in a broad range of growth conditions. Contrary the isolation of almost one order of magnitude smaller crystals of $Ce_2PdIn_8$ and $Ce_2PtIn_8$ has been more challenging. Their separation requires inspection of many samples by means of microprobe analysis in order to exclude inclusions of $Ce_3PdIn_{11}$ and $Ce_3PtIn_{11}$, respectively.

Specific heat data on $Ce_3PdIn_{11}$ and $Ce_3PtIn_{11}$ show a double peak structure pointing at a complex magnetic structure below 1.7 K and 2.2 K, respectively. The enhanced Sommerfeld coefficients support the heavy fermion character of both compounds. Further details of magnetic, transport and thermal properties of the compounds is a subject of intensive studies and will be published elsewhere [10].

**Acknowledgements**

This work was supported by the Czech Science Foundation (Project P203/12/1201). Experiments were performed in MLTL (http://mltl.eu/), which is supported within the program of Czech Research Infrastructures (project no. LM2011025). Crystallographic part was supported by Czech Science Foundation (project no. P204/11/0809).